\documentclass[spanish,english]{article}
\usepackage[T1]{fontenc}
\usepackage[latin9]{inputenc}
\usepackage{geometry}
\usepackage{amsmath}
\usepackage{amssymb}
\usepackage{stmaryrd}

\makeatletter

\newcommand*\LyXZeroWidthSpace{\hspace{0pt}}

\newenvironment{lyxlist}[1]
	{\begin{list}{}
		{\settowidth{\labelwidth}{#1}
		 \setlength{\leftmargin}{\labelwidth}
		 \addtolength{\leftmargin}{\labelsep}
		 }}
	{\end{list}}

\makeatother

\usepackage{babel}
\addto\shorthandsspanish{\spanishdeactivate{~<>}}

\begin{document}
\title{The Ontology and Semantics of Quantum Theory for Quantum Gravity}
\author{{\normalsize{}Alejandro Ascárate}\thanks{FaMAF-UNC (Córdoba, Argentina). Email: aleazk@gmail.com}}
\maketitle
\begin{abstract}
{\normalsize{}Based on a clear ontology of material individuals, we
analyze in detail the factual semantics of quantum theory, and argue
that the basic mathematical formalism of quantum theory is just okay
with (a certain form of ) realism and that it is perfectly applicable
to quantum gravity. This is basically a process about ``cleansing''
the formalism from semantic assumptions and physical referents that
it doesn't really need (we use the term ``semantics'' in the sense
of the factual semantics of a physical theory, and not in the sense
of model theory of abstract mathematics or logic). We base our study
on the usual non-Boolean lattice of projectors in a Hilbert space
and probability measures on it, to which we give a careful physical
interpretation using the mentioned tools in order to avoid the usual
problems posed by this task. At the end, we study a possible connection
with the theory of quantum duration and time proposed in \cite{key-4},
for which this paper serves as a philosophical basis, and argue for
our view that quantum gravity may show that what we perceive as change
in the classical world was just (an ontologically fundamental) quantum
collapse all along.}{\normalsize\par}
\end{abstract}
$\;$

\section*{{\large{}Introduction}}

$\;$

\selectlanguage{spanish}%
In the following, we will present some of the \emph{tools} developed
in \cite{key-1} to deal with the philosophical issues often found
in science. The goal is to formulate a purely physical, independent
and objective interpretation of quantum theory. Purely physical in
the sense that it will only refer to independent physical objects
and not to subjective or human elements (such as minds, knowledge,
uncertainty in knowledge, etc.). Independent and objective in the
sense that only genuine properties of the system are taken into account
and not elements that, although physical, are superfluous to it (such
as measuring devices, etc.).

\selectlanguage{english}%
$\;$

\selectlanguage{spanish}%
These tools are based on two basic pillars of philosophy: \emph{ontology}
and \emph{semantics}. In scientific theories, they are embedded in
them, although they are not always made explicit. The fact that scientific
theories use these tools should not be surprising. After all, one
of the central problems of ontology is to establish what the world
is really made of, which is also a problem of interest to science
(not to say equally central.) However, in science, ontology cannot
be arbitrary: it must always be compatible with materialistic realism,
which states that reality exists objectively and its fundamental substantial
components are only material entities. Our focus, however, will be
primarily on semantics; some additional ontological hypotheses will
be mentioned to the extent that they are necessary. 

\selectlanguage{english}%
$\;$

\section*{{\large{}1. Scientific Ontologies}}

$\;$

Science assumes that an external and independent Reality exists (its
famous method is just a consequence of this hypotheses; by the same
reason, one cannot use science's method to prove that ontological
hypotheses, one can just judge its correctness in terms of its fertility
and success.) Furthermore, Physical Theories assume that Reality \emph{is
made of something}. In ontology, this something is called a \emph{Substance}.
In science, it's assumed that Reality is made of a single type of
Substance, which is called \emph{Material Substance}. Therefore, science's
ontology is \emph{monistic}. In ordinary language, we refer to this
Substance simply as \emph{Matter}. Nevertheless, be aware that the
``matter'' of which we are talking about is not necessarily some
chunk of something with some mass, as, say, a piece of steel. The
notion of Matter used in science can become quite subtle and unintuitive.

$\;$

A key property of Material Substance is that two \emph{material individuals}
can \emph{Associate} to form a third material individual. If $a$
and $b$ are these two material individuals, we will denote the association
operation as $\dotplus$, and, in this sense, if $c$ is the third
material individual mentioned before, we write (we will denote as
$S$ the set which contains our material individuals):

\[
c=a\dotplus b.
\]

That is, in science we have different types of ``matter'', which
are each characterized by the binary association property, $\dotplus:S\times S\,\longrightarrow S$,
among its elements. Mathematically, we can postulate a Boolean $\sigma-$algebra
structure for the triple $(S,\dotplus,\boxempty$) (where $\boxempty$
represents the null element, or minimal element, $R$ the maximal
element, see below, and $\dotplus$ the ``disjunction'' or $sup$,
that is, the minimal upper bound between elements; one also has the
$inf$ or ``conjunction'', which is interpreted as the Interposition
$\dot{\times}$ of material individuals.) In physics, two typical
examples are the following:

$\;$
\begin{itemize}
\item if the elements of the set $S$ are pieces of charged material and
the association of two pieces to form a third is modeled by the operation
$\dotplus$, then the electric charge, $Q:S\,\longrightarrow\mathbb{R}^{+}$,
is \emph{additive} on $\dotplus$, that is, $Q(x\dotplus y)=Q(x)+Q(y),\,\forall x,y\in S$; 
\item if the elements of the set $S$ are static electric fields and the,
pointwise, superposition of two fields to form a third is modeled
by the operation $\dotplus$, then the intensity of the electric field
at point $p$ of space, $E_{p}:S\,\longrightarrow\mathbb{R}^{+}$,
is \emph{additive} on $\dotplus$, that is, $E_{p}(x\dotplus y)=E_{p}(x)+E_{p}(y),\,\forall x,y\in S$.
\end{itemize}
$\;$

In this way, we see how the proposed ontological theory is consistent
with scientific knowledge. In addition, we see how, in general, scientific
theories necessarily require some kind of ontological theory (of course,
the scientific theories themselves are the ones which suggest the
structure of the latter, although they do not usually determine them
completely.)

$\;$

The above is a non-trivial characteristic which we must postulate.
But, it also must be so that, when individuals associate, \emph{matter
doesn't disappear}: Material Substance is always \emph{conserved};
that is, if it exists, then it cannot suddenly disappear\footnote{The reader may be thinking in the supposed ``annihilation'' of matter
with antimatter, but that's not an annihilation, but rather a \emph{transformation}
of one type of material individuals (say, electrons and antielectrons)
into another type of material individuals (in this example, photons;
we can see how the naive picture of ``matter is mass'' can be quite
misleading.) Note that, in this example, conservation of matter is
related to conservation of energy (since the mass of the initial particles
is transformed into the light energy of the massless photons via $E=mc^{2}$);
we will come back to this latter.}.

$\;$

Other metaphysical theories may introduce other types of Substances,
like ``Immaterial Souls'', ``Mental Substance'' (this one is famously
associated with Descartes and his dualist, in terms of Substance,
solution to the, also famous, Mind-Body problem, i.e., what's the
mind and how does it interact with the physical body?), etc. But,
in science, \emph{only matter interacts and associates with matter}.

$\;$

All of this may seem a little too vague. Is there some way to characterize
matter in a more precise way? How do we know what is actually matter
and what cannot be considered as genuine matter? Fortunately, there's
a way. We postulate that all Material Substance carries \emph{Energy},
where this concept is defined according to the best physical theory
at disposition. In general, it's a Property (see below) of matter
which is represented, in one way or another, by an additive, real
function on matter (which depends on the reference frame adopted,
though), i.e., an $h(a)$ such that

\[
h(a\dotplus b)=h(a)+h(b),
\]
for any two elements $a,b$ from $S$, and which ``generates time
translations'' in the context of that theory (this is usually done
in what is called the ``Hamiltonian formulation'' of a dynamical
theory.) What this means is that \emph{Energy is the Property of a
Thing that allows us to determine, in a physical theory, how the Properties
of this Thing change in time} (see below for these terms.)

$\;$

Thus, to us, if something doesn't posses energy, then it cannot be
considered matter.

$\;$

Of course, the things that inhabit Reality are much more than mere
\emph{undifferentiated} bunches of material substance. The ``fauna''
of Reality is very rich, with individuals that can have very peculiar
and different \emph{Substantial Properties}. A Substantial Property
is a property/quality that a material element \emph{possess} or \emph{bears}
and it's as objective as the existence of the material element itself\footnote{This in contrast to Idealism, which states that properties are mental
constructs. As we will see below, in physical theories, we will \emph{represent}
properties by mental constructs, but we assume that the property exists
without us, and that, in fact, this representation may be only approximative.
Idealism identifies as a single thing the property and the mental
construct (although it accepts the existence of matter.)} (properties are not substantial elements themselves, though!) In
order to characterize a property and have a clear understanding of
its peculiarities, we need a physical theory about the material element
at issue. In this way, the property is characterized by \emph{all}
of the \emph{logical consequences} in the theory that involve this
property. For example, for a charged body, its electric charge is
one of its substantial properties and, in order to understand what
electric charge exactly is, we need to consider logical consequences
in a theory that involves the electric charge; for example, consider
the following logical consequence from classical electricity: ``two
bodies with charges of equal sign repel each other while two bodies
with charges of opposite sign attract each other''; evidently, this
proposition gives us valuable information that helps to elucidate
what the property ``charge'' is. We will call \emph{Thing} to a
matter element which possess substantial properties. Things also associate
to produce other Things. We denote the set of \emph{all} Things that
exist as $\Theta$. We mention that a very particular type of properties
are the so-called \emph{Emergent Properties}, that is, properties
which are present in the \emph{aggregate} Thing $c=a\dotplus b$,
but \emph{not} in its component Things, $a$ and $b$, before the
aggregation.

$\;$

Reality is the Thing $R\in\Theta$ which consists in the association
or aggregation of \emph{all} Things in $\Theta$.\footnote{One could say, what about time and space, are not they part of Reality
too? This is a complex issue. They are intrinsically tied to Things
in a certain specific sense, as explained in \cite{key-3}. Also,
what about logic? We use logic to understand Reality, but we don't
take it as part of Reality (it's not even a thing.)}

$\;$

\section*{{\large{}2. Scientific Semantics}}

$\;$

\subsection*{{\normalsize{}2.1 Reference}}

$\;$

\selectlanguage{spanish}%
A mathematical theory without a factual interpretation is only mathematical,
not physical. A scalar field satisfying the Laplace equation may be
the potential for a gravitational field or for a static electric field.
It is the factual interpretation (which interprets this field as the
potential for a gravitational field or for a static electric field)
the thing that transforms it into a theory of physics\footnote{Usually, the invention of some mathematical concepts is inspired by
the physical reality and are used to represent the elements behind
that inspiration; however, in many cases, the concepts prove to be
useful for representing other elements of physical reality, which
can be unrelated to the ones that inspired the invention of the concept;
a typical example is the concept of Hilbert space, which was originally
used to describe ordinary physical euclidean space, but later appeared
as the space of states of a quantum system and having infinite dimensions
instead of the usual three.}. And, of course, it is this map of interpretation (which maps mathematical
objects\footnote{These mathematical objects are mere mental \emph{constructs} made
by human brains. In this way, we do not adhere to mathematical Platonism
(which says they exist in their ``own objective reality''.) However,
this does not imply a free subjectivism, since different humans can
manage to understand each other through the same concepts. Then, in
practice, we can pretend as if Platonism were true. In addition, it
is assumed that classical Predicate Logic is applicable to both purely
mathematical predicates and to predicates with factual content. Hence
the utility of mathematics in the factual sciences, there's no mystery:
given a factual interpretation of the kind discussed here, we can
take advantage of the abstraction, precision, and systematic nature
of mathematics to study more accurately and systematically the physical
world (indeed, since we assume that classical logic applies to predicates
at both ends, then the conclusions derived in the purely mathematical
plane should in fact \emph{predict} or indicate a novelty in the factual
plane once the initial factual interpretation is applied to these
conclusions.)} to objects of the real world) the source of most of the problems,
due to the number of different, sometimes contradictory, positions
that are adopted in everyday practice and the literature.

$\;$

But we need the map to define and establish the theory, what do we
do then? How do we formulate a physical theory?

$\;$

There are two\footnote{There is a third type of semiotic interpretation called ``mathematical
interpretation'', in which abstract mathematical constructs are interpreted
in terms of other abstract mathematical constructs.} types of semiotic interpretations or semiotic maps:

$\;$
\begin{enumerate}
\item the purely semantic ($\varphi$); 
\item the pragmatic one ($\pi$). 
\end{enumerate}
$\;$

The purely semantic map, (1), assumes a certain kind of \emph{realism}
(which is an ontological hypothesis), simply the same that is needed
to give meaning to the fundamental maxim of the scientific method
(not to be confused with classical naive realism, in which \emph{all}
the properties of a given physical system always have to possess defined
values, etc.;

$\;$

\emph{}%
\begin{minipage}[t]{0.9\columnwidth}%
\emph{the realism we are considering here only assumes that there
are entities/things that exist and that have well defined and independent
physical properties; we stress: just properties, not necessarily a
definite value for ALL of these properties}; properties that have
a definite value are called \emph{manifest,} while, if they do not
have it, are called latent; furthermore, to be considered as existing,
a thing must have at least one manifest property; if a thing ceases
to have any manifest properties, then it ceases to exist according
to our view; we will come back to this in later.)%
\end{minipage}

\selectlanguage{english}%
$\;$

\selectlanguage{spanish}%
In this way, the interpretation here is very simple: factual item
$y$ of this physical system will be represented mathematically by,
say, the mathematical function $f$ (which is a construct.) And that's
all! We don't need anything else, we simply refer to the factual item
because it exists, it is simply there, we simply point to it (semantically
speaking, not pointing in a literal sense.)

$\;$

For example, if the physical entity is a particle and the property
is its classical electric charge, then at the abstract and mathematical
level we have a set $P$ and an additive function $Q:P\longrightarrow\mathbb{R}$;
the semantic map $\varphi$ relates these abstract mathematical elements
to objects of the real world:

$\;$
\begin{lyxlist}{00.00.0000}
\item [{i)$\:\varphi(P)=particles$;}]~
\item [{ii)$\:\varphi(Q(x))=electric\:charge\:of\:x\in P$.}]~
\end{lyxlist}
$\;$

Note that, in this kind of interpretation, if one introduces factual
elements which, however, lack a mathematical correlate in the theory,
then they are not true factual elements pertaining to this physical
theory and one can dispense with them. One should be careful to only
include in the factual base of the theory those elements for which
the semantic map is well defined and assigns a definite mathematical
correlate to them. Otherwise, we would be dealing with ``\emph{ghosts}'',
i.e., entities that are not modeled by the physical theory in consideration\footnote{Typical of the so-called ``instrumentalist'' and ``subjectivistic''
interpretations of quantum theory, which introduce things like ``observers'',
``apparatuses'', ``ignorance'', ``uncertainty'', and ``information'',
all things that have no actual mathematical correlate in the theory.}. Of course, these ghosts are just superfluous elements and should
be, as we said, cleared up from the factual base of the theory. \emph{This
is a key philosophical principle that one must always have in mind
when trying to provide a physical interpretation for the mathematical
formalism of quantum mechanics (or any other physical theory for that
matter), one really cannot overstate this point}. But this can be
very tricky if we don't have our physical theory precisely and clearly
formulated (which is, unfortunately, the exception rather than the
norm in today's physics.) This trickiness (together with the difficulty
in trying to measure directly the referents of a theory) is the reason
why most physicists tend to distrust ontology, particularly in matters
related to quantum theory; they prefer, instead, to resort to operationalist
approaches (which, ironically, actually makes matters even worse,
since this just opens the doors for ghosts to held a party inside
the theory!) Avoiding a problem doesn't solve it nor makes it disappear!

$\;$

The pragmatic interpretation, (2), does not assume an objective property
of the system; instead, it begins with a series of laboratory operations.
The interpretation would be as follows: the measurements made by these
particular laboratory operations will be represented mathematically
by, say, the function $f:\mathbb{R}\longrightarrow\mathbb{R}$. Notice
how drastically different these two interpretations are.

$\;$

For example, consider the color change on a litmus paper. We know
that this color change indicates acidity; but if we abstain only to
the spirit of (1), we cannot say that this change \emph{is} acidity
(as (2) would say.) Acidity is an objective property of the chemical
medium, while the color change of the litmus paper measures it.

$\;$

It should be noted that in the semiotic interpretations normally found
in science, not all the objects of the mathematical theory or structure
in question necessarily have a factual interpretation. This is called
a ``partially interpreted theory'' and is what is usually used in
scientific theories, which only make use of ``pieces'' of some mathematical
theory.

\selectlanguage{english}%
$\;$

\subsection*{{\normalsize{}2.2 Meaning}}

$\;$

\selectlanguage{spanish}%
When formulating a theory, it is only acceptable to use interpretations
of type (1). This is because we are only looking for one very basic
thing: basic physical meaning, \emph{anything else would be superfluous
here}. The formulation and interpretation of a theory is something
that belongs to the purely theoretical and conceptual plane (which
does not mean that reference is not made to factual items); a priori,
it has nothing to do with laboratories and laboratory operations.
Laboratories and laboratory operations become relevant in the stage
of empirical testing of theory, which comes later, \emph{once the
theory has been formulated}. When we make empirical tests of theories,
then it's there when we turn to interpretations of type (2). But,
in fact, the operations that one considers in (2) will depend on the
basic meaning of the concept according to (1). It is precisely the
meaning of the concept the thing that justifies, and allows us to
elucidate, the kind of operations that we need to consider.

$\;$

The word \emph{meaning} has been used; it is, therefore, convenient
to analyze it in more detail. The purely semantic interpretation takes
a mathematical construct and relates it to the real world. But this
is not enough to give physical meaning to this construct. The purely
semantic interpretation gives us, at best, the \emph{physical referents}
of the construct, that is, those physical \emph{entities} (i.e., the
Things that comprise the Reality hypothesized by the physical theory
in consideration) to which the physical interpretation refers. What
is important to note is that, in a full theory of physics, this construct
is not isolated. Indeed, there are also other mathematical constructs
in the theory, and, in general, they are all \emph{inter-related}
with one another (\emph{in the logical sense}.) These interrelations
constitute what is called the \emph{mathematical Sense} of the construct.
Some of these other constructs are also physically interpreted in
terms of purely semantic interpretations, and, thus, the mathematical
sense becomes also a \emph{physical sense}. In this way, \emph{given
a physical theory, we will adopt the point of view in which the meaning
of a construct is fully established by its total Sense and its Reference
class} ($\varSigma\ni\sigma$ and $\varSigma\subseteq\Theta_{Universe}$)
of Things ($\sigma$), \emph{both which can only be read once the
theory has been fully established in its mathematical axioms and semantic
interpretations}. This is the reason why we need to know all the logical
consequences that involve a Property in order to understand the Meaning
of that Property according to the physical theory in consideration.
Also note that, in a completely axiomatized theory, certain basic
constructs will determine the meaning of the other constructs of the
theory (in general, these basic constructs will be those whose physical
interpretations are made in terms of factual elements that are usually
taken as factual primitives, such as the notions of length, lapses
of time, facts, propensity, etc.)\footnote{Space and time, in particular, are quite generic to most theories.
In Scientific Ontology, one can actually make theories that give them
a \emph{precise} physical meaning. These theories are appropriately
called ``\emph{proto-physics}'' (see \cite{key-3} for a proto-physical
theory of spacetime.) Although, one often needs to insert them into
a physical theory in order to have a more \emph{exact} (in a \emph{quantitative}
sense) meaning.} It's only \emph{after} one has a clearly formulated theory that one
uses the meaning of its concepts according to it to design experiments
to measure them, and not the other way around, as in operationalism,
where semantics, meaning, and experimental testing of a theory are
all conflated and confounded.

$\;$

For example, the concept of mass in classical mechanics is mathematically
an additive function $M:P\longrightarrow\mathbb{R}^{+}$. The semantic
map $\varphi$ gives to it a physical interpretation as follows:

$\;$
\begin{lyxlist}{00.00.0000}
\item [{i)$\:\varphi(P)=particles$;}] where, also, $\varSigma=particles$;
\item [{ii)$\:\varphi(M(x))=inertia\:of\:x\in P$;}]~
\item [{iii)$\:M\;$appears}] in the equations of motion multiplying the
acceleration (which is another construct, with its own factual interpretation
and so on.)
\end{lyxlist}
$\;$

Clearly, i) and ii) concern the semantic interpretation part, while
iii) is related to the sense of the construct. Thus, it is the complete
theory of classical mechanics (its mathematical formalism, its semantic
interpretations, its laws, etc.) the thing which determines the meaning
of the construct in this example.

$\,$

It is well known that the three previous items in the mass meaning
are enough to propose, in a justified way, operational methods that
can allow the determination of the concrete numerical value of this
mass in the context of the empirical test stage of the theory. For
example, if $M$ is the value of the mass of a test particle and $M_{0}$
the value for a fixed particle taken as a reference for unity, we
know from the third law of motion that if the two particles interact,
then (in magnitude) $Ma_{M}=M_{0}a_{0}$. In this way, taking $M_{0}=1$
by convention, the value of $M$ can be calculated in terms of the
measured values of these accelerations through $M=a_{0}/a_{M}$ (this
is the supposed ``interpretation'' of the concept of mass given
by Mach; from our point of view, it's only one way of measuring the
value of mass, not a ``definition'' of mass, much less its physical
interpretation.) Thus, the meaning of mass is inertia, i.e., the same
force applied to a body of some mass gives it an acceleration which
is greater than the one it would give to a body with a mass greater
than the one of the initial one, i.e., inertia or mass is a ``resistance''
to being accelerated by a force. As mentioned, this is not in any
way a ``definition'' of mass, which is actually a \emph{specific
primitive} (i.e., beyond stating that it's a property and that it
appears in the mentioned law, it's left undefined) of this theory.
What we are doing is studying the logical consequences of the axioms
in order to understand what role this concept plays in the theory.
In Special Relativity, the meaning of mass, besides also being inertia
here, is expanded because of new features introduced by this theory,
like the one established in the famous equation $E=m_{0}c^{2}$, which
expresses that the (rest) mass, the inertia of a particle, can be
transformed into the energy (e.g., kinetic energy, electromagnetic
energy) of another particle, possibly massless (of course, these processes
are not allowed in classical mechanics, where the sum, $M_{o}=\sum_{i}m_{o}^{(i)}$,
of the total rest masses of the particles in a system is always conserved.)

$\;$

Of course, there may be many other operational methods, since this
is not the way to give meaning to the concept of mass, it is only
a way of calculating its concrete numerical value in a given experimental
context. It is in fact the well-defined meaning the thing that alone
can and should justify all these possible and different operational
methods. We have used the theory to justify the method proposed in
the example, which might suggest that there is some risk of circularity.
But this is not the case because the positive truth of physical theories
is not something that can be definitively proved and in terms of considerations
completely alien to it. Instead, or it is established that a theory
of physics is self-consistent and in accordance with observed facts
(of course, this will never be enough to ``prove'' the theory ``once
and for all'', this would fall into the severely criticized fallacy
of Empirical Induction), or it is established that it is inconsistent
and in disagreement with the observed facts. In this second case,
of course, the theory has been empirically falsified (in the Popperian
sense of the term; of course, this is also just a simplification of
a much more complex process in which theories are first amended many
times.) In this way, the method proposed in the example may actually
help to establish whether the theory is self-consistent in the empirical
context. In fact, if, once the values \LyXZeroWidthSpace \LyXZeroWidthSpace of
the masses have been obtained, the dynamics observed in a new experiment
does not agree with what is stipulated by the second law, where the
previous values \LyXZeroWidthSpace \LyXZeroWidthSpace measured for
the masses are used in this law, then something goes wrong in theory,
it is not a correct description of nature. If there is agreement,
then we simply obtain a transient verification of it (which, as we
have already said, is by no means a definitive proof of the theory),
and this is indeed the standard scientific practice.

$\;$

In a mathematical interpretation, the role of the referent is given
by the more specific mathematical construct. In general, this can
be precisely defined using mathematical axioms. In the case of the
factual interpretation, however, the referents are factual items,
which cannot be characterized by such definitions. The factual items
can only be pointed out or mentioned, not defined. In general, factual
items (entities, properties of entities, facts involving entities)
are characterized in a partial way by predicates, statements and propositions.
If these referents indeed exist in the physical reality, they can,
at best, be discovered in some experiment (although this is not necessary
for the formulation of the semantics of the physical theory in question;
not even its actual existence in the physical reality is needed, since
the physical theory and its referent are, in principle, just hypotheses
about the physical reality; concern about how to measure the existence
of referents at the formulation stage of the theory is unnecessary;
once the theory is empirically tested, it may in fact be proven that
the referents do not exist; however, this does not show that the theory
affirms its non-existence; in general, the theory needs them for its
formulation; what would happen if it were shown that the referents
do not exist is the falsification of the theory itself. All this should
not be surprising since, for example, when formulating theories such
as classical mechanics, the existence of its referents, classical
particles, is taken for granted; however, if we conform to the formalism
and standard interpretations of quantum mechanics, these referents
are pure fiction, there is no such thing as a classical particle in
the real world.)

\selectlanguage{english}%
$\;$

\subsection*{{\normalsize{}2.3 }\foreignlanguage{spanish}{{\normalsize{}Operationalism
in Modern Physics}}}

$\;$

\selectlanguage{spanish}%
In the operationalist philosophy of physics (like the usual ``Copenhagen
interpretation'' of quantum mechanics), the very formulation of the
theory is performed in terms of pragmatic interpretations of type
(2).

$\;$

This is very problematic due to the following reasons: in a pragmatic
interpretation, a large number of elements (such as measuring devices,
human observers, etc.) are usually brought to the scene and, although
they are necessary to obtain the concrete experimental value of a
property, they are completely alien and superfluous to what is really
necessary to simply give basic physical meaning to the property being
measured.

$\;$

As mentioned, this confuses two very different levels: the formulation
of theory (which is a purely conceptual and theoretical process) and
the stage of empirical testing of it (which should in principle be
a posteriori of the first.) This confusion may even introduce incoherence
into the theory, as exemplified by the Copenhagen interpretation of
quantum mechanics (according to which classical mechanics would be
among the hypotheses of quantum mechanics; this is absurd since quantum
mechanics is a theory which superceds classical mechanics.) The value
obtained in a pragmatic interpretation will, in general, depend on
the system being measured, of course, but also on the measuring devices
used and possibly on other details of the experimental context. This,
per se, is not problematic, provided it is clear that we are in the
stage of empirical testing of theory and not in its formulation.

$\;$

Consider a simple example. Let $\widehat{A}$ be a quantum operator
that represents some property $A$ of a quantum system. The standard
operational interpretation states the following: ``the eigenvalue
$a$ of operator $\widehat{A}$ is a possible result/value of measuring
$A$ in some given experimental context''.

$\;$

Now, let's analyze what quantum theory actually says (that is, let's
analyze how the theory is actually used by professionals in their
routine calculations.) Given that the theory provides an interpreted
mathematical formalism (in the semantic sense), then it is possible
to use an elementary tool of the semantics outlined here: identification
of the \emph{factual referents}. As mentioned above, when a mathematical
concept is interpreted in terms of physical elements, reference is
made to certain physical entities; these entities are the factual
referents. A quantum particle can be a referent, as well as a measuring
device and even the ``mind'' of a human observer.

$\;$

In this way, let's take the operator $\widehat{A}$, as it is usually
given by quantum theory, and let's investigate its factual referents
(to make the analysis more concrete, the operator in question could
be taken, for example, as the quantum z-spin; the Hamiltonian can
contain the environment in its variables, so it is an example to be
taken as a separate case.) And this is where the inadequacy of the
usual operational interpretations is evident: in the standard formalism
of quantum mechanics, both $\widehat{A}$ and its eigenvalues are
\emph{only} functions (in the semantic sense) of the quantum particle;
no reference is made to measuring devices (much less minds), nor is
the mathematical form of the operator dependent on such elements.
In more precise terms, let $P$ be the ``observable'' object of
study, $\varSigma\ni\sigma$ a factual reference class and $\varSigma'\ni\sigma'$
the class of physical environments or ambients of the elements of
$\varSigma$. If $P$ is a real and objective property of its referent
$\sigma\in\varSigma$, then we denote $P=P(\sigma)$. Operational
interpretations often state that, for any quantum property $P$, it
is generally given that $P=P(\sigma,\sigma')$, where $\sigma'\in\varSigma'$
is the environment of referent $\sigma$. However, this is usually
the exception rather than the rule. If we take the example of z-spin,
standard quantum mechanics states that the z-Pauli matrix continues
to represent the z-spin, regardless of which environment the quantum
particle is immersed in: the theory itself states that the same matrix
represents this property for both the free (Hamiltonian only kinematic)
particle and the particle in interaction with, for example, a field.
In view of examples such as these, it is clear that it is capricious
to insist that the referent of quantum mechanics is an irreducible
``object-apparatus'' (or even ``object-mind'') entity rather than
physical systems (and their corresponding objective properties) that
can, in principle, be considered independently (at least ideally)
from their surroundings.

$\;$

Thus, we know that the referent has to be just the system (and not
a system-apparatus, system-instrument, system-observer, system-mind,
etc., complex) because one can see explicitly in the mathematical
description of the properties in this theory that there's no room
for the environment in them. Let's take another example. For a particle,
interactions with external entities are introduced in the theory by
adding a ``potential energy'' term $\widehat{V}$ to the kinetic
one. The total energy one gets is, then: $\widehat{H}=\frac{\widehat{p}^{2}}{2m}+\widehat{V}$.
By studying the formalism, one can see that the mathematical form
of $\widehat{p}=-i\hbar\frac{\partial}{\partial x_{j}}$ (a property
of the particle) is left unchanged if we change the type of interaction
characterized by $\widehat{V}$ (where one has to put apparatuses,
instruments, observers, etc.; ``ideal measurements/devices'', which
wouldn't appear here, are non-sense, physically.) This can \emph{only}
be the case if $\widehat{p}$ is a property of the quantum system
alone, i.e., a property that exists objectively and not something
that arises via the interaction with an instrument (as operationalists,
and perhaps also radical relationalists, believe.) The same analysis
is valid for other properties. Now, when there's an interaction, evidently,
the probabilities will change. But, for example, for a particle, the
theory says that there can indeed be free particles (just take $\widehat{V}=0$),
and that probabilities in the theory still make sense in that case\footnote{As an analogy, consider the position $x_{t}(\sigma;K)$ of a particle
$\sigma$ moving under the influence of a spring of spring constant
$K$. Thus, of course, the explicit form of $x_{t}(\sigma;K)$ will
depend on the environment. Indeed: $x_{t}(\sigma;K)=x_{0}\,\mathrm{sin}\left(\left(\sqrt{\frac{K}{m(\sigma)}}\right)t\right)$.
But this is only an effect of the dynamics; if $K=0$, there's no
force, no spring, the particle is free, but $x_{t}(\sigma)$ \emph{still
remains there}, \emph{this means that the position was actually a
property of the particle in the first place}. The same is valid for
the probability $P_{t}(\sigma)$ of a free quantum particle $\sigma$
(see next section.)} (of course, they do not depend on any environment, since there's
none; the typical ones are the ``wave-packets'', which are probability
distributions that center around, e.g., a position value and only
depend on that number, the dispersion $\Delta x$ around it, and time.)
This can only be the case if probabilities in quantum theory refer
to an actual Property of the system alone in the first place, and
not something else (for example, if they were mere ``information''
that some observer has about the system, then some variable there
should say\emph{ which} observer); in particular, if, for a free particle,
the probability is $1$ for some value of a property, then the only
possible interpretation (i.e., one that doesn't introduce superfluous
elements lacking correlates in the variables) is that the value is
simply being possessed by the particle, in the same way in which a
classical particle possess a value, even when there's nothing measuring
it (this point of view assuming that the description given by quantum
physics is complete.) In both cases, for probabilities values and
for the values of other properties, it's the mathematical formalism
the one that suggests which interpretation suits it and which doesn't;
in these cases, we saw that it suggests realist, objective and non-operational
interpretations. All of these semantic considerations are necessary
since we want to know what's exactly the referent of this theory.
Furthermore, the \emph{only} way to shed light to this matter is by
a semantic analysis, as the one we just did

$\;$

We can see, then, that the operational interpretations only manage
to attach to $A$ useless and unnecessary referents, which make no
difference in the predictions made by the concrete theory. When this
happens, the interpretation is said to be adventitious. Most pragmatic
interpretations make this conceptual mistake massively when they are
used to give a supposed physical meaning to some construct; they introduce
physical entities (like the physical environment of the particle,
in the previous example) that remain, however, \emph{orphaned from
a mathematical correlate in the theory}; that is, entities that are
thus transformed into phantoms, which are smuggled into theory (and
removed with equal ease when it is convenient, since they make no
real difference.) Then, in the spirit of (1), we rephrase as: ``the
eigenvalue $a$ of the operator $\widehat{A}$ is a possible value
of the objective property $A$ (of the referred system, of course)
represented by this operator''. That is, reference is made only to
the quantum system and its objective properties.

$\;$

Note that, in principle, it would be possible to develope a theory
in which the operator $\widehat{A}$ in fact depends on the measuring
apparatus; moreover, further research could prove experimentally that
this new theory is superior to standard quantum mechanics from the
empirical point of view. However, this new theory would be just that:
a new theory (superior and distinct from standard quantum mechanics.)
What the previous semantic analysis shows is that, as far as the formalism
of standard quantum mechanics concerns, there are no references to
measuring apparatuses in the operator $\widehat{A}$ or its eigenvalues. 

$\;$

In this way, we see that semantic interpretations (and, in fact, the
whole theory) are actually also hypotheses about reality, which could
eventually prove to be merely approximations to the latter. However,
theories must be formulated at some point, which implies making these
hypotheses explicit. In the case of quantum mechanics, this conveys
to make explicit the hypothesis that establishes that the properties
of the system will not be considered as dependent on its surroundings.
Perhaps this is not correct, perhaps it is only an idealization of
reality, but this is work for a new theory; these are two distinct
levels that should not be confused. The concern about how to measure
the existence of referents in the formulation stage of the theory
is unnecessary; once the theory is empirically tested, this can in
fact prove that the referents are non-existent; however, this does
not show that the theory affirms its nonexistencen general, the theory
needs them for its formulation; what would happen if it is shown that
the referents do not exist is the falsation of the theory itself.\foreignlanguage{english}{
All this should not be surprising since, for example, in formulating
theories such as classical mechanics, the existence of their referents
(classical particles) is taken for granted; however, if we adapt to
the formalism and standard interpretations of quantum mechanics, these
referents are pure fiction, there is no such thing as a classical
particle in the real world.}

$\;$

The above considerations do not, however, mean that the environment
(considered as a physical entity that exists objectively and in which
we include, for example, measuring devices, human observers, a photon
gas, etc.) cannot interact with a quantum system: it can, but the
form of this interaction is something that must be modeled by the
quantum theory itself, that is, it is not a part of the formulation
of the latter; instead, the corresponding, and possibly not trivial,
temporal evolution of the probabilities (for example, they may become
more acute distributions on a given value after interaction with the
apparatus, the surrounding environment, etc.) is something to be explained
by the theory once it has been formulated and the environment included
in its Hamiltonian. In fact, the way in which the physical environment
of a quantum system can cause the sharpening of probabilistic distributions
of the latter is a topic of debate and research in contemporary physics
(i.e., the need for an adequate measurement \emph{theory}.)

\selectlanguage{english}%
$\;$

\section*{{\large{}3. }\foreignlanguage{spanish}{{\large{}(Objective) Probability
and Quantum Theory}}}

$\;$

\selectlanguage{spanish}%
Let us consider the \emph{facts, $h$}\footnote{These facts usually\emph{ involve some property of the system}, e.g.,
a possible fact can be that ``the position of the system is $x=5\,m$''.
In the case of completely classical theories, one usually adopts the
point of view according to which the values of properties are always
defined. However, results like the Kocher-Specker theorem, establish
that this is impossible in quantum physics (at least for \emph{all}
properties \emph{at} the same time.)}\emph{, that can happen to a thing} (we take ``happening'' as something
dichotomic: a fact either happens or not, we don't accept ``partial
happening'', whatever that could be.) We include under this notion
the facts that, although they can be in principle possible and accessible
for this concrete thing, never happen to the system in the real world;
facts that actually happen are called ``real'', while the rest are
called ``potentials'' (in this way, real facts are a proper subset
of potentials); given the potential facts, $h_{1}$ and $h_{2}$,
the disjunctive fact ``$h_{1}$ or $h_{2}$'' only makes sense as
a potential fact and not as real, since only conjunctive facts of
the type ``$h_{1}$ and $h_{2}$'' (that is, the fact in which both
facts occur simultaneously) can actually occur in a given system;
thus, disjunction only makes sense as potential, not as reality; when
we speak of the ``set of facts'', in general we will be referring
to the total set of potential facts (note that these \emph{physical}
conjunctions and disjunctions of \emph{facts} must, and should, always
be Boolean operations, in order to make any sense and match intuition.)
This is so because physical probability theories always operate at
the level of potential facts; in particular, to include such a disjunction
of potential facts. The distinction between potential and actual facts
becomes relevant when one wants to empirically contrast the theory,
since it only makes sense to empirically test situations related to
real facts.

$\;$

In this way, the facts that can occur to the system are elements that
are also capable of appearing as the corresponding factual objects
in a possible semantic interpretation. This is the case in physical
probability theories, where, physically, probability provides the
instantaneous \emph{intensity} of the \emph{propensity} of a \emph{single}
system towards the occurrence of a given fact; this is a \emph{key}
insight by, among others, Popper into the whole debate about probabilities
and their interpretations. That is, for us, the propensity, the mere
potentiality, of a system towards the possible occurrence of a given
fact that involves it, is a real, \emph{manifest, actual}, instantaneous
and objective \emph{property} of it. This categorization is, of course,
a hypothesis of the ontological type. 

\selectlanguage{english}%
$\;$

\selectlanguage{spanish}%
Now, given a theory and its respective laws, the temporal evolution
of probability will dictate which facts are more likely than others
as the system evolves; in analogy to mechanics, we could say that
there is a ``dynamic'' part, an evolution ``constrained by the
laws of motion'' (that is, the laws of physical theory), while the
set of all potential facts and all probability measures are the ``kinematic''
part of a theory of probabilities. In general, one first must consider
the system and all of the physical properties that the physical theory
attributes to it; if it's a quantum field, these could be the quantum
energy or momentum of the field, etc. Thus, the potential facts for
this system will all be occurrences which involve the eventual definition/acquisition
of values for some properties (or that this value is in some subset
of the real numbers.) For example, consider a spin particle $\frac{1}{2}$.
The latter implies that given a direction $\overrightarrow{n}\in\mathbb{R}^{3}$,
the projection of the spin of the particle in that direction can only
take the values $S_{\overrightarrow{n}}=\pm\frac{1}{2}$. Then, the
facts $h_{\pm\frac{1}{2}}$ in which the value of this projection
is indeed $\pm\frac{1}{2}$ are possible facts for this particle and
therefore fall into our definition of potential facts (note that it
will be the dynamics of theory the one that will say if the particle
indeed acquires some of these values or not.) On the other hand, the
alleged fact $h_{\frac{1}{3}}$, in which the value of the projection
is $S_{\overrightarrow{n}}=\frac{1}{3}$, will never happen; but,
this, not because the fact is a potential one and the dynamics dictamines
its non-occurrence, but, rather, because, being the particle of spin
$\frac{1}{2}$, it is not even in its intrinsic potentiality to acquire
that value for the projection; that is, the fact $h_{\frac{1}{3}}$
is not a potential fact for this particular system under consideration.

$\;$

In the experimental testing stage of the theory, it is evident that,
given this interpretation of probability, the \emph{experimental frequency}
obtained by measuring many identical systems will approximate the
theoretical value of probability (this is due to the law of large
numbers\footnote{Which is a mathematical theorem derived from the mathematical machinery
of probability theories (we stress, it's neither a definition nor
an interpretation of probabilities, it's just a theorem) that, in
simple situations (such as the throw of a dice), can be physically
interpreted in the sense that, if one makes $N$ identical trials,
and where the dice has the same intensity of the propensity for a
given fact in each of them, then the number $n$ of identical trials
in which the fact actually happens increases when the mentioned intensity
increases, which is what we would expect. Even more, as the number
of identical trials gets very large, then the frequency $\frac{n}{N}$
approximates better and better the theoretical value for the intensity
of the propensity.}.) However, we emphasize that this frequency cannot be taken as the
meaning of probability; instead, it is simply a particular way of
measuring its value in the laboratory; for example, another method,
other than the above, often used in standard experimental practice
is to measure the intensities of spectral lines in a spectroscopic
apparatus. Clearly, these two distinct methods are simultaneously
justified only by adopting the objective probability interpretation,
and for a single system, mentioned in this section.

\selectlanguage{english}%
$\;$

\selectlanguage{spanish}%
More precisely, given a physical system and a notion or pre-theoretical
conception of some of its properties\footnote{As is the previous notion of propensity; it is said pre-theoretical
in the sense that we will define a physical theory as a mathematical
formalism interpreted through factual semantics and then the previous
physical conception is only halfway, still orphaned of a mathematical
formalism.}, we can \emph{exactify} and \emph{elucidate} it (and thus construct
a physical theory) if we can find a given mathematical structure $M$
(\emph{exactification}) such that some of its elements can be interpreted
factually (\emph{elucidation}) in terms of the physical notion under
consideration. For example, for the case of probability, a physical
theory of probability is obtained by specifying elements $m$ y $m'$
en $M$ such that the semantic map $\varphi:M\longrightarrow factual\,items$
acts on them as follows:

$\;$
\begin{lyxlist}{00.00.0000}
\item [{$\qquad\varphi(m)=fact$$\quad$and$\quad\varphi(m')=intensity\:of\:the\:propensity\:of\:this\:fact$.}]~
\end{lyxlist}
$\;$

We'll call the pair $(M,\varphi)$ (and in which factual referents
and their properties are implicit) an exactification-elucidation and
possible physical theory of probabilities. Thus, the ``exactification''
of the concept of intensity of propensity is, once given this notion,
the process of finding this semantic map and mathematical theory $M$
for a given system (or class of systems.)

$\;$

Of course, there may be many different mathematical structures that
allow us to do this, which will predict different behaviors for probability.
In the definitions below, we will take the pre-theoretical concept
of probability as intensity of propensity and look for different exactifications
for it. In particular, we will see that what fundamentally differentiates
quantum theories from classical ones is precisely that they exactificate
this notion of probability with different mathematical structures.
This simple distinction will be, for us, the starting point of quantum
theories. 

$\;$

\selectlanguage{english}%
$\mathbf{Definition\,3.1}$ \foreignlanguage{spanish}{(\emph{Physical}
Theory of Classical Probabilities) Let }$(X,\Sigma(X)$$,\mu)$\foreignlanguage{spanish}{
be an abstract theory of classical probabilities}\footnote{\selectlanguage{spanish}%
A well known \emph{purely mathematical} result that will be mentioned
later is the following. \foreignlanguage{english}{(\emph{Weak Law
of Large Numbers}) Let $\left(X,\Sigma(X),\mu\right)$ be an abstract
theory of probabilities, $\left\{ f_{n}\right\} _{n\in\mathbb{N}}$
an infinite sequence of measurable functions $f_{n}:X\,\longrightarrow\mathbb{R}$,
and such that their probability distributions are identical and independent
(and for which $<f_{n}>_{\mu}$ and $Var(f_{n})\leq\sigma^{2}$ exist
for all functions, that is, $<f_{n}>_{\mu}<\infty$ and $\sigma^{2}<\infty$;
thus, it is obvious that $<f_{n}>_{\mu}=M,\,\forall n\in\mathbb{N}$
, where $M$ is some fixed positive real number.) Then:}

\selectlanguage{english}%
\[
Pr-lim_{n\rightarrow\infty}\overline{f_{n}}=M,\,\overline{f_{n}}\doteq\frac{1}{n}\sum_{k=1}^{n}f_{k}.
\]
}\foreignlanguage{spanish}{ (i.e., a measure space $X$ with measure
$\mu$ such that $\mu(X)=1$). Consider now a physical system, the
set $H$ of all the potential facts that involve it and the intensities,
$I_{h}$, where $h\in H$, of the propensities to occur of these facts.
Also, given two facts $h_{1}$ and $h_{2}$, we will denote as $h_{1}\land_{f}h_{2}$
to the fact that consists in their physical conjuntion, and as $h_{1}\lor_{f}h_{2}$
to the (exclusively potential) fact that consists on their physical
disjunction. Then, if one has a factual semantic interpretation of
the previous abstract theory (or an exactification-elucidation of
the physical concepts in terms of this abstract theory) such that
the semantic map $\varphi$ acts as:}

\selectlanguage{spanish}%
$\;$
\begin{lyxlist}{00.00.0000}
\item [{$\varphi(E_{h})=h$}] (with $E_{h}$ arbitrary element of $\Sigma(X)$
and $h$ its corresponding fact in $H$),
\item [{$\varphi\left(\mu(E_{h})\right)=I_{h}$,}]~
\end{lyxlist}
$\varphi(E_{h_{1}}\bigcap E_{h_{2}})=h_{1}\land_{f}h_{2}$ and $\varphi(E_{h_{1}}\bigcup E_{h_{2}})=h_{1}\lor_{f}h_{2}$\footnote{Note how we obtain the intuitive notion of the addition of probabilities
for mutually exclusive (i.e., \foreignlanguage{english}{$E_{h_{1}}\bigcap E_{h_{2}}=\textrm{Ø}$)}
facts: $I_{h_{1}\lor_{f}h_{2}}=\varphi\left(\mu(E_{h_{1}}\bigcup E_{h_{2}})\right)=\varphi\left(\mu(E_{h_{1}})+\mu(E_{h_{2}})\right)$.} for all pair of elements $E_{h_{1}},E_{h_{2}}\in\Sigma(X)$,

$\;$

we call the pair $\left[\left(X,\Sigma(X),\mu\right),\varphi\right]$
a Physical Theory of Classical Probabilities\footnote{While this is itself a definition, it contains elements that are actually
physical hypotheses, such as the existence of a semantic map that
acts in a given way, etc.}$.\;\square$

$\;$

\selectlanguage{english}%
Given the notion of this last definition, two comments come to place:

$\;$
\begin{itemize}
\item First, \foreignlanguage{spanish}{it is only in the context of physical
probability theories that the usual names of ``event space'' for
the measure space and ``random variables'' for measurable functions
are consistent; in a purely abstract mathematical theory, they are
completely out of place.}
\item Second, it is now possible to see the powerful physical implications
of the law of large numbers. Suppose we have a physical system described
mathematically by a classical theory of probabilities and we want
to \emph{measure experimentally} the numerical value of some average
value $<f>_{\mu}$ (of a property of the system represented by the
function $f$) to \emph{compare it with the value predicted by the
}theory, in order to make an \emph{empirical contrastation} of the
latter. A priori, the physical theory does not suggest any experimental
method that we can apply to measure this value. However, the law of
large numbers offers us another mathematical way to calculate this
value (namely, in terms of ``frequencies'' of certain infinite sequences.)
\emph{Given the physical interpretation of probability} as intensity
of the propensity, what is important is that this \emph{does} suggest
an operational method (one of many possible) to obtain an \emph{empirical
estimation} of the value at issue and which consists simply in taking
a very large number of identical physical systems, and prepared in
the same initial state (since, when interpreted physically, the condition
$<f_{n}>_{\mu}=M$ means that all the systems must have the same propensity
and probability distribution for the property being measured, which
translates into them all having the same initial state; note that
the use of different functions $f_{n}$ for the same property rather
than a single one is just an artifact to model the \emph{different}
values of the outcome of the measurement, what is relevant here is
just the probability distribution for these values), in measuring
the value\footnote{That is, we assume here that the system is classical and that always
has defined values. In this way, by measuring it, one simply \emph{reveals}
the value that the property already possessed. The situation is not
so clear and simple for quantum systems; this is discussed in detail
later on.} of the property at issue, and finally in calculating the value of
the frequency and \emph{empirical averages} obtained from the measurements
in question. According to the law of large numbers\footnote{Actually, the theorem doesn't say anything about \emph{finite} numbers
of experiments, it only mentions the \emph{infinite} limit. Thus,
we are extracting here from it more than what it says. More refined
theorems, that tackle this subtle issue, exist. Here, we just mention
the basic reasoning here with this version of the theorem, even if
it's not strictly correct, in order to illustrate what's the actual
relation between probabilities and experimental frequences, which
are often confounded (more on this later).}, the larger the number of identical experiments, the empirical values
should be closer and closer to the theoretical value for the average
value of the property predicted by the theory (if the latter is in
fact correct, of course.) That is, it is the physical theory of probabilities
(complete mathematical formalism plus its semantic interpretation)
the thing that justifies the operational methods to put it to empirical
test. As explained before, this is typical of theories in which their
concepts have well-defined physical \emph{meanings}.
\end{itemize}
\selectlanguage{spanish}%
$\;$

\selectlanguage{english}%
$\mathbf{Definition\,3.2}$ \foreignlanguage{spanish}{(Quantum Probability
Measure) \cite{key-3} Let $(\mathcal{H},(\cdot,\cdot))$ be a complex
Hilbert space and $\mathfrak{P}(\mathcal{H})$ its lattice of orthogonal
projectors. A quantum probability measure over $\mathfrak{P}(\mathcal{H})$
is a function $\rho:\mathfrak{P}(\mathcal{H})\,\longrightarrow\left[0,1\right]$
such that:}

\selectlanguage{spanish}%
$\;$

a) $\rho(\mathbf{0})=0$ and $\rho(\mathbf{1})=1;$

$\,$

b) $\rho\left(\vee_{k\in\mathbb{N}}P_{k}\right)=\sum_{k=1}^{\infty}\rho(P_{k})$
if $P_{k}\in\mathfrak{P}(\mathcal{H}),\,\forall k\in\mathbb{N},$
and $P_{k}P_{j}=\mathbf{0}$ when $k\neq j.\;\square$

$\;$

It is simple to prove that, under the assumptions of (b), $\vee_{k\in\mathbb{N}}P_{k}=s-\sum_{k=1}^{\infty}P_{k}$.

$\;$

Note that, on each maximal set of commuting projectors, $\mathfrak{P}_{0}(\mathcal{H})\subset\mathfrak{P}(\mathcal{H})$,
the \emph{restriction} of $\rho$ to any of them is simply the equivalent
of a classical probability measure in $\mathfrak{P}_{0}(\mathcal{H})$.
However, by switching to a different $\mathfrak{P}_{0}(\mathcal{H})$,
in general, this restriction results in a classical measure which
is different from the one arising from the restriction on the first
subset considered. In this way, we can see the theory of the classical
measure as contained and also generalized in our previous definition.

$\;$

$\mathbf{Definition\,3.3}$ (\emph{Abstract} Theory of Quantum Probabilities)
We will call abstract theory of quantum probabilities to the triple
$\left(\mathcal{H},\mathfrak{P}(\mathcal{H}),\rho\right)$, where
$\mathcal{H}$ is a complex Hilbert space and $\rho$ a quantum probability
measure\footnote{By Gleason's theorem, \foreignlanguage{english}{\emph{there is a positive,
trace-class operator, of trace} $1$, $T$, \emph{(determined completely
by $\rho$) such that} $\rho(P)=tr\left(TP\right),\,\forall P\in\mathfrak{P}(\mathcal{H}).$
As a corollary (Kochen-Specker), then $\rho$ \emph{cannot} take values
\emph{only} in $\left\{ 0,1\right\} \subset\left[0,1\right]$.}} on $\mathfrak{P}(\mathcal{H}).\;\square$

$\;$

In line with the previous comment, we can see that one of the fundamental
characteristics of quantum probability theories is that they introduce
a ``new mathematical degree of freedom'' (absent in the classical
case), namely the notion that two elements $P,Q\in\mathfrak{P}(\mathcal{H})$
\emph{can commute among them or not. }As we began to see, this brings
important consequences, like what was said in the previous comment.
This motivates the following definition:

$\;$

$\mathbf{Definition\,3.4}$ (Compatible and Incompatible Elements)
Let $(\mathcal{H},(\cdot,\cdot))$ be a complex Hilbert space and
$\mathfrak{P}(\mathcal{H})$ its lattice of orthogonal projectors.
Two elements $P,Q\in\mathfrak{P}(\mathcal{H})$ \emph{are called compatible
whenever they commute with each other, that is,} $PQ=QP$. Two elements
$P,Q\in\mathfrak{P}(\mathcal{H})$ \emph{are called incompatible if
they do not commute with each other, that is}, $PQ\neq QP.\;\square$

\selectlanguage{english}%
$\;$

\selectlanguage{spanish}%
$\mathbf{Definition\,3.5}$ (\emph{Physical} Theory of Quantum Probabilities)
Let $(\mathcal{H},\mathfrak{P}(\mathcal{H})$$,\rho)$ be an abstract
theory of quantum probabilities. Consider now a physical system, the
set $H$ of all the potential facts that involve it and the intensities,
$I_{h}$, where $h\in H$, of the propensities to occur of these facts.
Given two \emph{certain} facts (i.e., not just any two given facts;
this will become clear below) $h_{1}$ and $h_{2}$, we denote as
$h_{1}\land_{f}h_{2}$ to the fact that consists in their physical
conjunction, and as $h_{1}\lor_{f}h_{2}$ to the (exclusively potential)
fact that consists on their physical disjunction. Then, if one has
a factual semantic interpretation of the previous abstract theory
(or an exactification-elucidation of the physical concepts in terms
of this abstract theory) such that the semantic map $\varphi$ acts
as:

$\;$
\begin{lyxlist}{00.00.0000}
\item [{$\varphi(P_{h})=h$}] (with $P_{h}$ arbitrary element of $\mathfrak{P}(\mathcal{H})$
and $h$ its corresponding fact in $H$; furthermore, the correspondence
in question must be a bijection),
\item [{$\varphi\left(\rho(P_{h})\right)=I_{h}$,}]~
\end{lyxlist}
$\varphi(P_{h_{1}}\land P_{h_{2}})=h_{1}\land_{f}h_{2}$ and $\varphi(P_{h_{1}}\lor P_{h_{2}})=h_{1}\lor_{f}h_{2}$
for all pair of \emph{compatible} elements $P_{h_{1}},P_{h_{2}}\in\mathfrak{P}(\mathcal{H})$,

$\;$

we call the pair $\left[\left(\mathcal{H},\mathfrak{P}(\mathcal{H}),\rho\right),\varphi\right]$
a Physical Theory of Quantum Probabilities\footnote{In this way, on each $\mathfrak{P}_{0}(\mathcal{H})\subset\mathfrak{P}(\mathcal{H})$,
the physical theories of quantum probabilities are reduced to a classical
physical theory of probabilities.}$.\;\square$

\selectlanguage{english}%
$\;$

Given this last definition, two comments come to place:

$\;$
\begin{itemize}
\item First, when describing \emph{different} physical systems by means
of physical theories of quantum probability, we are talking about
different theories, because their concrete referent is different,
even if the background is always a physical theory of quantum probability.
This is why it would be more appropriate to talk about quantum theories,
in the plural, instead of a quantum theory. In fact, what one does
is to take this notion of physical theory of quantum probability,
apply it to different types of concrete systems (namely, particles,
fields, etc.), obtain the corresponding physical theories and study
their particular properties. Note that we were relatively vague when
talking about what types of facts are contained in $H$. In principle,
we will consider that this set contains all the possible facts that
can happen to the physical system in consideration (in particular,
if it has a physical property valued in the real numbers, the facts
in which the value of this property is defined and contained in some
$B\in\mathcal{B}(\mathbb{R})$.) These facts will depend on which
are the particular properties of the concrete system under consideration;
in this way, the facts alluded to refer to the physical entity as
it is modeled and described by the theory (for example, if the referent
is a quantum field, possible facts can be those referred to properties
such as the energy or momentum of the field, both properties introduced
by the theory itself); we can see, then, that theories introduce their
referent and its properties; in this way, its existence in reality
is hypothetical (note that this hypothetical existence is something
that has no relation to the use of the concept of potential fact,
which was used in the previous definition, as something that can potentially
occur to the referent once it is hypothesized that it exists.) In
other cases, we will not stop to try to build or make explicit $H$
for each particular system, but we will assume that $\left[\left(\mathcal{H},\mathfrak{P}(\mathcal{H}),\rho\right),\varphi\right]$
exists; we will do this since we are interested in studying quantum
theories in a general way; that is, the object of study will almost
always be a generic quantum theory. We will refer to the factual referent
$\sigma$ of a generic \emph{Physical} Theory of Quantum Probabilities
as a \emph{Quantum System}.
\end{itemize}
$\;$
\selectlanguage{spanish}%
\begin{itemize}
\item \emph{We will not bother in trying to define (beyond the pre-theoretical
realm)} $h_{1}\lor_{f}h_{2}$ or $h_{1}\land_{f}h_{2}$ when $P_{h_{1}},P_{h_{2}}\in\mathfrak{P}(\mathcal{H})$
are \emph{incompatible }elements (and, thus, much less the action
of the semantic map), since, as we will argue later in Proposition
3.2 below, \emph{these are not valid potential facts for a quantum
system}; also, we will not give a factual interpretation to the lattice
operations $P_{h_{1}}\land P_{h_{2}}$ or $P_{h_{1}}\lor P_{h_{2}}$
when $P_{h_{1}},P_{h_{2}}$ are mutually incompatible (although, the
projectors $P_{h_{1}}\land P_{h_{2}}$ or $P_{h_{1}}\lor P_{h_{2}}$
themselves can indeed represent some fact, what we are saying is that
we don't give a physical interpretation, in terms of conjunction or
disjunctions of the facts $h_{1}$ and $h_{2}$, to the lattice operations,
in this case; in this way, the non-Boolean lattice we use is \emph{partially
interpreted}.)
\end{itemize}
\selectlanguage{english}%
$\;$

We will discuss later the so-called measurement problem, the role
of the law of large numbers in quantum probabilities, and other related
topics. We don't discuss them here since \emph{they are not part of
the basic semantic axioms of quantum probabilities.}

$\;$

\selectlanguage{spanish}%
$\mathbf{Proposition\,3.1}$ (Characterization of Incompatible Events)
Let $\rho$ be a quantum probability measure and $P$, an event such
that $\rho(P)=1$. Then, if $Q$ is \emph{(purely}\footnote{That is, it cannot be further decomposed as the sum of a noncommuting
\emph{and a commuting} part with $P$.}\emph{) incompatible} with $P$, it always results that $0<\rho(Q)<1$,
where the inequality is strict$.\;\square$

$\;$

In this way, we can see why the equivalent of global Dirac measures
in quantum probability theories do not exist: as the previous proposition
shows, it is because of the existence of incompatible events. This
result, valid for any quantum probability measure, is not possible
in classical probability theories (both because the existence of global
Dirac measures and by the fact that quantum measures always distinguish
when events are incompatible, which is remarkable since in the definition
of the latter only properties of the projectors intervene, they have
nothing to do with the measures) and shows us the radical mathematical
and physical consequences of the existence of incompatible elements/events
in theories of quantum probability.

\selectlanguage{english}%
$\;$

Now, a key result that one can rigorously prove only in the framework
we have adopted (or a similar one), since it's a result that involves
not only the mathematical apparatus of the theory, but also its physical
\emph{semantical} interpretation and the elements related to it. To
most physicists, it's obvious; nevertheless, it's rarely stated clearly
so that one can identify all of the very different elements that intervene.

$\;$

\selectlanguage{spanish}%
$\mathbf{Proposition\,3.2}$ There is \emph{no} projector $P_{h_{1}\land_{f}h_{2}}\in\mathfrak{P}(\mathcal{H})$
such that $\varphi(P_{h_{1}\land_{f}h_{2}})=h_{1}\land_{f}h_{2}$
for (supposed) potential facts $h_{1}\land_{f}h_{2}$ given by the
conjunction of two incompatible facts, $h_{1}$ and $h_{2}$. That
is, a physical theory of quantum probabilities does not admit, or
cannot model, the mentioned facts, and, therefore, \emph{they are
not} in the first's factual base.

$\;$

\emph{Proof}: Consider $h_{1,2}=h_{1}\land_{f}h_{2}$, defined at
the pre-theoretical level as a fact which happens \emph{iff both $h_{1}$}
and $h_{2}$ happen simultaneously. Now, in classical probability,
given an event $E_{h}\neq\mathbf{0}$ that models the potential fact
$h$, there's a Dirac measure $\delta_{s},\,s\in E_{h},$ which is
such that $\delta_{s}(E_{h})=1$; in quantum probability, given an
event $P_{h}\in\mathfrak{P}(\mathcal{H})$, $P_{h}\neq\mathbf{0}$,
that models the potential fact $h$, the projector can always be decomposed
as $P_{h}=\sum_{i=1}^{\infty}u_{i}(u_{i},\cdot)$, where $\left\{ u_{i}\right\} _{i=1}^{\infty}$
is an orthonormal basis of the projection space, and then there is
a $T_{h}=\sum_{i=1}^{\infty}C_{i}u_{i}(u_{i},\cdot)$, where $\left\{ C_{i}\right\} _{i=1}^{\infty}$
is any series such that $\sum_{i=1}^{\infty}C_{i}=1$ (for example,
the geometric series $\left\{ \left(\frac{1}{2}\right)^{i}\right\} _{i=1}^{\infty}$),
such that $\rho_{T_{h}}(P_{h})=tr\,\left(T_{h}P_{h}\right)=1$.\footnote{Indeed, intuitively, if the fact $h$ is genuinely in the potentiality
of the system (and, thus, actually modeled by the theory), that is,
it can indeed happen, then its eventual occurrence (although the dynamics
may establish that it does not happen) has to be able to be described
by the theory, just to be in agreement with this. For this particular
theory, if an event occurs, then the state of the system must be such
that, at the instant that this happens, it assigns probability equal
to $1$ to this fact, since it is actually occurring (note that we
are not saying that this probability has to be equal to 1 before the
fact occurs.) Thus, if a fact can happen, there must be at least one
state in the theory in which, if the system is in that state, then
this fact happens.} Suppose $h_{1,2}$ is modeled by $P_{h_{1,2}}\in\mathfrak{P}(\mathcal{H})$,
$P_{h_{1,2}}\neq\mathbf{0}$ (thus, it's indeed in the potentiality
of the system); then, there are states of the system, say $\rho_{h_{1,2}}$,
which give a probability equal to $1$ to this fact. Suppose the system
is in one of these states. Then $h_{1,2}$ has maximum intensity in
its propensity to happen. By our definition of $h_{1,2}$, this, in
turn, requires that both facts, $h_{1}$ and $h_{2}$, must also have
maximum intensity of their propensity to happen, i.e., probability
equal to $1$, each of them, on that same state $\rho_{h_{1,2}}$.
However, as we saw earlier in Proposition 3.1, there is no such state
for pairs of incompatible facts. Thus, any state $\rho_{h_{1,2}}$,
whose supposed existence implied this contradiction in the first place,
cannot actually exist, and this implies that the fact $h_{1,2}$ cannot
be modeled by some $P_{h}\in\mathfrak{P}(\mathcal{H})$, $P_{h}\neq\mathbf{0}$,
since, if this were the case, then these states have to exist, as
we mentioned above. Furthermore, it cannot be modeled by $P_{h_{1,2}}=\mathbf{0}$
either. Indeed, we have $P_{h_{1}}\neq\mathbf{0}$ and $P_{h_{2}}\neq\mathbf{0}$
(otherwise, they would commute), and $\rho(P_{h_{1,2}})=0$, for any
measure. Since $P_{h_{1}}\neq\mathbf{0}$, there must be a measure
$\rho_{h_{1}}$ such that $\rho_{h_{1}}(P_{h_{1}})=1$. Suppose the
system is in that state, then $\rho_{h_{1}}(P_{h_{2}})=0,$ since
$\rho_{h_{1}}(P_{h_{1,2}})=0$ and the definition of $h_{1,2}$, but
this contradicts Proposition 6. Thus, the fact $h_{1,2}$ cannot be
modeled by the theory$.\,\square$ 

$\;$

An analogous result holds for the disjunction $h_{1}\lor_{f}h_{2}$
of incompatible facts. In this way,\emph{ since a possible factual
element but lacking a mathematical correlate in the formalism of the
theory cannot be considered as a genuine factual element belonging
to the whole of which the theory makes semantic allusion, we conclude
that it is not in the intrinsic potentiality of a quantum system the
possibility of the occurrence of the alleged facts $h_{1}\land_{f}h_{2}$
or $h_{1}\lor_{f}h_{2}$ when $h_{1}$ and $h_{2}$ are incompatible
with each other}.

\selectlanguage{english}%
$\;$

We can see that this result is not a consequence of the non-distributivity
of the lattice alone, but more a consequence of the whole formalism
of quantum theory itself, no matter how we frame it:\emph{ in a quantum
system, these facts simply do not have a place and we must just live
with that.}

$\;$

\selectlanguage{spanish}%
In relation to the material above, we make some pertinent comments
here. The fact that the lattice $\mathfrak{P}(\mathcal{H})$ is non-distributive
has brought, throughout the history of the concept (that goes back
to the years 1930s with von Neumann), diverse and quite extravagant
interpretations, all of them incorrect, to us, and based on (to our
point of view) elementary semantic misunderstandings. The most widespread,
and old, of these is to denominate the physical theories of quantum
probability as ``Quantum Logic'' and to make the interpretation
that the non-distributivity of the lattice implies in some way a ``new
non-distributive logic for reality''. 

\selectlanguage{english}%
$\;$

\selectlanguage{spanish}%
First, note that we have completely ruled out the usual practice of
naming the facts $h$ as ``possible experimental propositions about
the system'' (\emph{this is a key distinction we make here}, i.e.,
propositions vs. facts) and the probability as ``probability that
the proposition is true in a experiment''. This is so because, in
physics, one speaks of the \emph{probability of a fact}, not of a
``proposition''; in addition, bringing up the concept of ``truth''
entangles our vocabulary even further: the notion that a theory of
physics is true or not is something that is established, in general,
when comparing its predictions with the experiment, it has absolutely
nothing to do with the (purely conceptual) stage of its formulation;
this latter comment also applies if one takes the notion of ``experiment''
used in these propositions as belonging to some kind of operationalist
physical interpretation. Facts and propositions are two different
concepts. For example, the fact that the position of the system is
defined and is $q=5m$ is something other than the proposition $P=$``The
position of the system is $q=5m$''; in the first case, we are talking
about a factual item that inhabits and occurs in the physical reality
(i.e., those things that physical theories aim to describe), while
the second is a mere statement, \emph{which can admit different types
of truth} (Theoretical truth, if compared with theoretical statements
derived from a theory; Factual truth, if compared with statements
obtained from empirical results\footnote{Even more, the concept of factual truth applies to propositions and
may be different from the value of probability, according to theory,
of the associated fact; for example, a theory can predict the value
$1.0$ for the probability of $h$ and an experiment may determine,
however, that the (factual) truth value of the corresponding statement,
``the fact $h$ occurs for sure'', is, for example, false.}\footnote{\selectlanguage{english}%
A given proposition can be true according to one body of knowledge
(either empirical or theoretical), but false according to another.
This point of view is particularly adequate in the sciences, where
one first makes deductions from hypotheses (which we take provisonally
as true) to later put these deductions to some test; if the test says
the deductions afirm something which is not true (in relation to some
body of knowledge), then we correct the hypotheses, not the deductions
(assuming, of course, that they are logically correct.)

$\;$

In particular, we don't take the point of view in which \emph{all}
propositions are ``born'' with an ``intrinsic, definite, objective
and eternal'' truth value; instead, the given body of knowledge in
consideration will establish if some proposition has a truth value
and which one it is, i.e., it will \emph{assign}, \emph{if possible},
a truth value to the proposition. Compare this with the platonist
view on logic, in which all propositions are born with an intrinsic
and unalterable truth value (to the point in which a proposition and
the statement that says that this proposition is true are basically
taken to be the same thing) and where one can always state things
like ``(p and q) is true iff (p is true) and (q is true)'' (for
this, the truth value of p and also q must, of course, be defined
from the beginning, as in the platonist view.)\selectlanguage{spanish}%
}.) Physical theories of probability apply to the first, since they
seek to mathematically model factual items; no physicist speaks of
``the probability of a statement'': s/he speaks of the probability
that a fact will happen. 

\selectlanguage{english}%
$\;$

\selectlanguage{spanish}%
The set of propositions (which includes statements derived from/about
facts $h$, not the fact itself; note that a fact has always associated
at least one statement, the one it says that the fact occurs, while
the converse is not necessarily valid) is in fact a Boolean algebra
under the operations of logical conjunction and disjunction of statements,
but this structure inhabits the abstract plane and in fact logic itself
does, it's not even passible of being refuted empirically since it
does not refer to any physical system in the first place! (note that
logic is about connectives and propositions and it's completely independent
from the propostions' truth values.\footnote{The interpretation of these connectives is just the standard, logical
one, which is a purely conceptual notion and, in particular, they
don't necessarily have any relation to the conjunction and disjunction
of physical facts about a physical system (this possible relation,
if any, should be established by the semantic interpretations of the
physical theory.)}) In fact, the interpretation of a Boolean lattice as the logic of
propositions is a typical example of what we called an abstract or
mathematical interpretation. The fact that a Boolean lattice can be
used to model propositional logic and also the set of facts in a classical
classical theory of probability only shows the flexibility and vast
applicability of the abstract mathematical concept of Boolean lattice.
As should be obvious, in all the demonstrations and arguments one
does in quantum physics, one always uses the ordinary (Boolean) usual
propositional logic.

\selectlanguage{english}%
$\;$

\selectlanguage{spanish}%
In other words, if $e(h)$ represents the propositional statement
of fact $h$ (i.e., $e(h)="h\,happens"$), \emph{the logical conjunction
and disjunction of $e(h_{1})$ and $e(h_{2})$ are always defined
and are Boolean operations for every pair of facts in both classical
and quantum. }If in physics we then decide to use another lattice
in our physical theory of probabilities, this implies exactly nothing
in relation to ordinary logic\emph{. }If $P_{h_{1}},P_{h_{2}}\in\mathfrak{P}(\mathcal{H})$
are compatible, then we will postulate the point of view in which
the propositional statement of $h_{1}\land_{f}h_{2}$ (respectively,
$h_{1}\lor_{f}h_{2}$)\emph{ is given by the logical conjunction of
$e(h_{1})$ and $e(h_{2})$ }(respectively, disjunction)\footnote{For the case of compatible elements, we know that they are part of
Boolean sublattices of $\mathfrak{P}(\mathcal{H})$, and thus the
homomorphism of lattices that we are establishing between these Boolean
sublattices and the Boolean algebra of propositions is correct and
well defined. Also, since we don't give to the connectives of $\mathfrak{P}(\mathcal{H})$
an interpretation as connectives of facts when they are incompatible,
we don't have then a homomorphism between this full non-distributive
lattice and the set of facts, with conjunctions/disjunctions of incompatible
facts included (the connectives on this set, we recall, should always
be distributive, since we are dealing with their known intuitive physical
meaning). We could also identify the propositional statement of $h_{1}\land_{f}h_{2}$
(respectively, $h_{1}\lor_{f}h_{2}$) with the logical conjunction
of $e(h_{1})$ and $e(h_{2})$\emph{ }(respectively, disjunction)
even for incompatible facts, but this will remain only at the pre-theoretical
level since, as mentioned, the facts $h_{1}\land_{f}h_{2}$ (respectively,
$h_{1}\lor_{f}h_{2}$) are not valid potential facts for a quantum
system.}. 

\selectlanguage{english}%
$\;$

\selectlanguage{spanish}%
Thus, with this and the results of the main text, we can see then
that the value of theoretical truth (with respect to the quantum formalism
developed so far) of the proposition or statement that affirms $e(h_{1})\land_{L}e(h_{2})$,
with $h_{1}$ and $h_{2}$ incompatible with each other, is \emph{always
defined} and is simply \emph{always false}.

\selectlanguage{english}%
$\;$

\selectlanguage{spanish}%
Note that we only claim that the proposition $e(h_{1})\land_{L}e(h_{2})$
is false (because the fact $h_{1}\land_{f}h_{2}$ is not in the \emph{potentiality}
of the system, since it's assumed that $\mathfrak{P}(\mathcal{H})$
exhausts all the potential events, and thus it can never happen) but
\emph{not} that this truth value has been obtained by analysing the
(probability, i.e., with respect to a given state or probability measure)
truth values of $e(h_{1})$ and $e(h_{2})$ separately for this case.
This because, since $h_{1}\land_{f}h_{2}$ is not modeled by the theory,
the probability truth of $e(h_{1}\land_{f}h_{2})\equiv e(h_{1})\land_{L}e(h_{2})$
is not defined and thus trying to analize it and calculate it via
the standard rules would be as inconclusive as trying to find a state
$\rho$ in classical mechanics such that for the fact $h_{3}$, in
which the position of the particle is $x=3$, gives a probability
$\rho(h_{3})=0$ for a particle that can only move in the interval
$\left[0,2\right]$: this cannot happen since $h_{3}$ is not in the
domain of any probability measure (which act on potential facts) because
$h_{3}$ is not a valid potential fact for this particle ($x=3$ is
not in the kinematics of the particle.) Also note that the conclusion
we reached here was done by analysing both the mathematical formalism
and its physical interpretation; in particular, the meaning of the
notion that the fact $h_{1}\land_{f}h_{2}$ is not in the potentiality
of the system, since $\mathfrak{P}(\mathcal{H})$ exhausts all the
potential events, and thus it can never happen.

\selectlanguage{english}%
$\;$

\section*{{\large{}4. Is the Reality of Quantum Theory Fundamental?}}

$\;$

Note that physical probability theories, such as those in the previous
section, only attribute to its referent new (i.e., besides the ones
already considered and which depend on the nature of the system) properties
such as the propensity, nothing more and nothing less. That is, the
propensity, the mere potentiality, of a system for the occurrence
of a given fact that involves it, is a real, actual, and objective
property of it. What is real, is perfectly defined and exists at all
times is the quantum system itself and its different propensities;
this quantum system is the bearer of these propensities.

$\;$

The question that names this section is actually rather tricky and
rivers of ink have flowed in trying to address it. From one side,
it's indeed a fundamental theory, in the sense that its laws and predictions
have not been superseded by a new and more precise theory. The debate
revolves, instead, around the question about what is exactly the underlying
reality ``behind the surface'' of a quantum probability theory,
in a sense which we will explain in the next paragraph. 

$\;$

In a classical probability theory, there are probability measures
$\delta$ (called Dirac measures) such that, for \emph{any} fact they
assign to it a probability of, strictly, either $0$ or $1$, that
is, absolute certainty that the fact occurs or that it doesn't. The
fact could be, e.g., about a given value of energy, and these measures
will tell us with precision if that's actually the value of energy
or not. In this way, \emph{all} properties of the system have a well
defined value. Thus, we can forget about probability, just take the
trajectory $x(t)$, which we know is well defined, and build our mechanics
around it. In this sense, if we have a probability here, we could
say that, while which values occur and which do not occur was already
intrinsically clear, for some reason or another we cannot describe
the system with precision and thus we have to conform ourselves with
a crude approximative theory with a referent only having a propensity
for these values\footnote{That is, we have two theories here: one whose referent only has propensities
(of intensity less than one) and another whose referent has a definite
trajectory. The first one is an approximation of the second one. In
particular, from the point of view of the first theory, the propensities
are true real properties of its referent, and not ``information''.
From the point of view of the second theory, the first theory is,
of course, incomplete, and this is often loosely stated by saying
that the propensities of the first theory are just ``(incomplete)
information that the experimenter has about the system'' or that
the probabilities are just ``epistemic''; the correct interpretation
is the one we mentioned and not these last ones, there's only one
type of probabilities (interpreted physically as propensities, which
are considered real properties of the system; thus, it's not a question
of semantics), the epistemic part is in the approximation process
of one theory by another.} (determinism and defined values, though, are still still there, they
are just behind the ``surface''; thanks to this, in experimental
situations, one can interpret the law of large numbers in a way similar
to the example of the throw of a dice.) When the theory is compatible
with an underlying reality of this type (where all the properties
have defined values), one says that the theory admits ``hidden variables''
(them being, of course, the ones that make all the properties to have
defined values.) Note that the ``ignorance of information'' take
on probabilities, where they are not considered as real properties
of the system, and, instead, after an experiment, one simply checks
the system to ``update'' our information about its definite values,
seems to suggest an underlying hidden variable reality (this is its
true content.)

$\;$

Now, the issue with quantum probability theories is that probability
measures of the type of $\delta$ cannot exist now! (this key result
is the Kochen-Specker theorem.) This means we can't do here the trick
that we did in the classical case, that is, a full, completely objective
underlying hidden variable reality is simply not possible in quantum
physics (there's still a possibility for a restricted type, see discussion
about the measurement problem below). Thus, at \emph{any} given situation,
there will be facts for which the intensity of their propensity to
happen is greater than $0$ but less than $1$\footnote{This is caused by the existence of incompatible facts in quantum physics:
indeed, it's imposible for two incompatible facts to have, each one,
an intensity of their propensity equal to $1$, and, even more, for
some type of incompatible facts, the one which has probability less
than $1$ cannot have probabilty equal to zero, either.}. In this way, we seem forced to conclude that, if the value of a
property is associated to one of these facts, it will be undefined.
That is, \emph{not all properties of the system will have defined
values}, i.e., the very core of classical physics turns to be invalid.
Now probabilities become the primary variable in the theory rather
than, e.g., trajectories. In particular, it also means that the conjunction,
$h_{1}\land h_{2}$, of two incompatible facts, $h_{1}$ and $h_{2}$,
is \emph{not} an admissible fact for a quantum system, i.e., its eventual
occurrence is not allowed even in principle (since, if the occurrence
of their conjunction were permisible, this eventual occurence would
mean the simultaneuous happening of the two component incompatible
facts, something which the quantum formalism doesn't allow even by
a trick like the one of the classical case, because Dirac measures
cannot exist on sets of facts which include facts incompatible with
each other; thus, if the value of a property associated with one of
these facts is defined, then, the value of a property associated with
the other, is not, and viceversa.)

$\;$

Properties that have a definite value are called manifest, while those
which don't, are called latent\footnote{Based on classical intuition, one may think that this distinction
doesn't seem to make much sense, since, if a property is not manifest,
then one would think it's not there, that it doesn't exist. But, in
classical physics, we can consider that the time evolution of all
the defined values is given by a time dependent Dirac measure $\delta_{t}$;
thus, the difference in quantum physics is that the dynamics is given
by a time dependent general measure $\mu_{t}$. With this we can see
that there are actually two layers, one is the fixed set of all the
possible values that a property can take (which can be discrete),
and other is the time dependent probability measure which indicates
what's the propensity for these values. It's in the measure where
the interactions with other things is included and this is the way
in which they interfere with the dynamics. Nevertheless, all the machinery
of the property is actually there, before this second level, and,
as we mentioned, is represented by a variable that is only associated
to the quantum system alone. This is why we say that the particle
can take or not a defined value for a property rather than saying
that it ``takes a property'', as a radical \emph{relationist} would
say. As discussed below, the interaction with apparatuses makes a
particle, perhaps via the dynamics (in the usual view on the issue),
to take a defined value for one of its (latent) properties, the property
itself is not ``created'' by this.}; furthermore, to be considered as existing, a thing must have at
least one manifest property; if a thing ceases to have \emph{any}
manifest properties, then it ceases to exist according to our view.
Some say that the implicance, in quantum physics, that not all properties
can be manifest at once implies the ``death of realism in physics''.
But this is a naive classical view of realism. The fact that a system
can have some property whose value is not defined does not imply that
the system ``does not exist'', the system always exists and always
has certain real and actual \emph{manifest} properties defined: the
very \emph{propensities} of the system for the occurrence of facts.
The position value it can adopt is not a reliable indicator of its
continuous existence as in classical mechanics.

$\;$

Nevertheless, the view in which values are not defined implies an
evident problem: when and how these systems acquire a definite value,
then? There are some cases in which this is very clear and non-controversial
(for example, an electron in a homogeneous magnetic field, along the
$z-$axis, is such that the value of the $x-$component of its spin
periodically oscillates back and forth between being defined and undefined;
this is a very simple exact solution to the equations for the time
evolution of a quantum system.) But, there are some other cases which
seem less straightforward to analyze. Indeed, the physical evidence
seems to suggest the following. Given two Things $x$ and $y_{Q\mid R}$
($R$ being a property of $y_{Q\mid R}$), if Thing $x$ has Propensities
for the different values of a Property $Q$ of it, then the aggregation
with $y_{Q\mid R}$, to form $x\dotplus y_{Q\mid R}$, will make the
fact $h_{q}$, for some possible value $q$ of $Q$, to happen at
some point, but one cannot predict with previous information \emph{which}
particular value will take place: that will depend on the Intensity
of the Propensity towards each value that the system had, but it only
depends on the value for this intensity that it had at the very moment
before the aggregation, that is, the aggregation doesn't modify the
system's intrinsic tendency, it only makes the fact $h_{q}$ to happen
in an apparently unpredictable way. Thus, at the end of the aggregation,
the situation becomes similar to the application of probabilities
in the classical case: \emph{one has a definite value}, but only a
probability to predict which. In physical experiments, where the previous
behaviour\footnote{Even when assuming this is needed to measure quantum systems, it's
not a semantical assumption here (it may be in operationalist approaches,
where probability is semantically interpreted in terms of experimental
frequences, but this is untenable since, as we mentioned, probabilities
in quantum physics are only a function of the system alone, and, also,
frequences only appear mentioned in a theorem derived from the basic
mathematical machinery of probability theory, it plays no role in
its axioms), instead, it's just a physical assumption or hypothesis
that it's done when measuring quantum systems, and one which is supported
by the evidence. It's needed to give a usable physical interpretation
to the law of large numbers, but it's not mandatory for the basic
formalism (with its semantic assumptions) to necessarily give this
theorem a physical interpretation, if not by the mere fact, as we
have seen here, that, unlike the dice example, it's not a trivial
thing to do. These situations are common in physical theories (for
example, one cannot measure temperature with a gas thermometer without
some additional physical information, not contained in the theory
of thermodynamics, about the behaviour of this gas.)} is observed, the law of large numbers (\foreignlanguage{spanish}{which
is also valid in quantum mechanics, since every fact is contained
in a Boolean sublattice of the lattice of projectors}; in this case,
the sublattice is the one for which $P_{h_{q}}\in\mathfrak{P}_{0}^{h_{q}}(\mathcal{H})$)
is then used to get an empirical estimate of the probabilty, which
is compared with the theoretical value, and this is the standard procedure
to test the predictions of quantum physics. The problem of definite
outcomes in this type of interactions is often called or known as
``the measurement problem'', because of its role in experimental
measurements of quantum systems (it's also called the ``collapse
of the probability function'' problem, since, in these processes,
the function seems to pass abruptly from a value which is less than
$1$ to the value $1$; although, this may be an incorrect view, since,
at the microscopic level, the value may be still undefined and what
makes it to seem defined is the interaction together with an idealized
classical interpretation of the result.) Others think that the problem
is much more far reaching and that it consists in the very problem
about how quantum indefiniteness reconciles with the obvious definiteness
of the classical, macroscopic world. In any case, the problem is rather
complex and admits, at the very least, two layers: i) it's rather
unclear in what exactly consists the full and complex interaction
process that makes a quantum system acquire a definite value (or appear
to do it) when interacting with another physical system (this other
system can be, but not necessarily, a measurement apparatus; the key
aspect, however, seems to be that this second system is usually a
\emph{macroscopic} one) and if it can be explained by quantum physics
itself (including modeling the apparatus itself as a quantum system,
too) or another supplementary theory is needed (those who believe
that quantum physics, as it's known today, is the fundamental theory
and that no hidden variables or modifications to the current formalism
are needed, are naturally inclined to the first approach, which is
rather problematic because the way in which quantum systems evolve
in time seems precisely to forbid processes of this kind, but, on
the other hand, this also is relative to if one is considering that
the system actually acquires the value or just appears to do so from
a classical, macroscopic perspective; the problem is often stated
as a trilemma in which at least one of the following three options
always has to be rejected in order to avoid a logical contradiction
in the known mathematical formalism of quantum physics: a) there are
no hidden variables, b) measurements always yield single, defined
outcomes, c) the standard time evolution of quantum physics holds...
the problem is that all the three options seem reasonable and difficult
to give up! this is the root of all the controversies around the problem,
since any way out from it implies the sacrifice of a cherished belief;
thus, in particular, if one doesn't want hidden variables, see paragraph
below, and believes that the interaction implies a quantum dynamical
process in which the probability distribution becomes sharp, then
it seems a modification of the standard quantum dynamics will be needed);
ii) its central role in the very procedure with which quantum physics'
predictions are experimentally tested. The ``measurement problem''
is one of the most infamous, controversial and discussed aspects in
the foundations of quantum physics (it has been around, without an
universally accepted solution, since the very origin of quantum physics
itself, in the late 1920s; both Einstein and Schroedinger famously
condemned the standard interpretations of quantum physics because
of it; the latter with his famous experiment in which the ``value
of life'', which can be either dead or alive, of a cat, a macroscopic
thing, is undefined if described by quantum physics alone.) It should
also be noted that, while there seems to be indeed some sort of indeterminism
(since one can't predict \emph{which} value a property is going to
take because we only have an intensity for the propensity towards
the occurrence of each of these values), it's still unclear if quantum
systems acquire a manifest property \emph{only} after an interaction
or if they can \emph{spontaneously} do so\footnote{For example, in the radioactive decay of radioactive materials, the
nucleus of an atom seems to spontaneously split into two lighter nuclei,
and, of course, while emitting radiation in the process.} (which is an even more radical form of indeterminism; although, one
may consider that an interaction with the ever present, but diffuse,
environment causes it.)

$\;$

We mention, though, that there's actually a final attempt to give
some classical flavor, i.e., hidden variables, to quantum physics.
As we mentioned, classical probability is locally valid over sets
of facts related to properties that can take definite values simultaneously,
that is, compatible facts. Thus, at each of these local sets, we have
local Dirac measures and, then, we can do the trick we did in the
classical case: for the properties involved in the local set, all
values are well defined and the probability given by quantum physics
is just an approximation. But, we can only do this at only one set
at a time, since a same property may belong to two different local
sets but the values assigned to it in each of them cannot be, at least
for one property, the same (since, otherwise, we risk at getting a
global Dirac measure.) Thus, this, by itself, is not very helpful.
The move is to bring back operationalism, to some extent, and say
that a set of measuring devices/apparatuses (or, maybe, a very particular
type of environment, just to avoid talking about instruments in an
abstract way and the usual conceptual fog that this brings) prepared
to measure the properties for a given local set ``select/create''
one of these local valuations for it. Now, as we already saw, the
mathematical form of an observable (and, thus, its possible values)
doesn't depend, in the quantum formalism, on the ``context'', and,
then, the role of the measuring instruments, the context, is just
to create (each time we measure) the valuations which give different
values each time and that we cannot predict with the quantum formalism
alone (presumably, which, exactly, are the actual assigned values
will depend on the details of the instruments' composition; note that
this environment still can appear in the total energy function $H$
of the quantum theory and, in this way, may still produce at least
some effects in the dynamics that can be described by it.) As a first
point, the reader may already have noticed the similarity bewteen
this and what we called the measurement problem, it's still here,
just disguised in another form: while values are technically defined,
when we pass to another context, the value assigned to a same property
changes abruptly, or, properties incompatible with the ones of the
previous context, suddently acquire values; the difference is that,
here, the explanation of how this happens is outside the quantum formalism,
while in the standard collapse approach it's assumed that the collapse
process should be explained by the formalism, since its reality is
assumed fundamental. Having said all this, if we restrict to situations
in which there's always some measuring setup around the quantum system,
then the difference between the two different approaches in discussion
reduces to: either, 1) the measuring context creates a valuation (via
some ``contextual hidden variables'', i.e., hidden variables that
change from context to context), which we don't know with precision,
and the quantum formalism just serves to give a crude probabilistic
approximation, or 2) the measuring context makes the propensity to
go sharp, or to ``collapse'', at some values, which we don't know
with precision, for the different compatible properties of that context
and where the occurence of a given value depends on the intensity
of the propensity at the instant before the measurement (intensity
whose value is exactly equal to the probability value of the previous
point.) A thing to note is that, in the second approach, if the probability
is $1$ for some fact, then one interprets that it's happening, since
the probability measure provides a complete description of the system;
but, in the first approach, it can only be considered to be happening
if measuring devices are giving a valuation to the context to which
it belongs (if they are measuring another context, then the values
associated to the first one are undefined, since the valuation is
now in the second context; thus, the probability provided by quantum
physics should be interpreted with care here, since the interpretation
in which they are complete descriptions will give a misleading picture
of the underlying reality\footnote{Note that, unlike the above, in the classical case, where the (objective)
valuation is just a single and global one, if the probability is one,
the valuation also agrees (for this value) with the picture given
by the approximative theory.}.) Another thing to notice with the first approach (besides the fact
that it, still, cannot assign, simultaneously, definite values to
incompatible properties, since their contexts are different) is that
it's not valid for a completely isolated particle, which is a widely
used notion in, e.g., particle physics and perturbative QFT. A further
questionable aspect of these interpretations is the completely ad
hoc nature of the hidden variables (in fact, most of what happens
would actually be described by something outside quantum physics,
the latter being reduced to a rather modest role) and their rather
peculiar ``contextual'' dependence, which is anathema to classical
physics, that is, precisely the thing that one supposedly wanted to
recover in the first place with all this approach; thus, the victory
is at best a pyrrhic one.

$\;$

Thus, quantum theory can be interpreted either with contextual hidden
variables or with propensity as fundamental (that is, without an underlying
hidden variable reality.) That is, in either case, if we want interpretations
with a coherent semantics, the result is a realist ontology, since
at every moment at least one property has a defined value (in the
first, given by the ``context'' properties used to define the contextual
hidden variables, and, in the second, simply the propensity itself.)
Furthermore, in both cases, the so-called wavefunction collapse can
be added as a \emph{non-semantic axiom} (unlike the operational interpretations,
where the collapse is indeed a semantic axiom, and, therefore, the
theory cannot even be formulated without postulating it, nor can the
theory be used to explain it; in the realist approaches, is not a
semantic axiom but it can be postulated as a law if necessary, and,
in this way, it still can intervene in the meaning of, e.g., probability,
the law of large numbers, and experimental frequencies.) It should
be stressed, though, that in whatever of the two cases, the interpretation
of probability in quantum theory must be done in terms of propensity
if one pretends semantic coherence (of course, in the hidden variables
case, it's not going to be the fundamental reality, but only an effective
one.)

$\;$

\section*{{\large{}5. Quantum Gravity: a Possible Way Out}}

$\;$

Thus, regarding collapse, we have, as discussed in the previous section,
two options with two sub-options each. That is, when a certain \emph{context-defining
interaction} occurs, then, either the quantum probability distribution
(taken as fundamental and objective) becomes sharp, or the values
of some contextual hidden-variables define (for this particular context,
and get undefined for incompatible contexts). Furthermore, this sharpening/value-definition
is either modeled as some explicit time evolution or dynamics (in
the first case, necessarily non-linear; in the second, some dynamical
equations for the hidden values must be provided), or, in both cases,
it's added as an ad-hoc postulate without further clarification (of
course, this breaks the usual Schrödinger evolution by definition).

$\;$

Now, a possible \cite{key-4} phenomenon in quantum gravity may be
the discretization of time duration into a succession of finite fundamental
processes. Furthermore, each of these processes is subjected to the
usual quantum randomness in their possible occurrence. That is, a
context-defining interaction will randomly produce these ocurrences.
Physically, for this picture to be coherent, it must be such that
time advances as a succession of instants which are very close to
each other in proper time distance and in which the duration of the
instants themselves is very small. Thus, at the macroscopic scale,
this is perceived as a succession of instants, each of duration zero,
which forms a continuum whose subsets have finite duration, and which
monotonically increases (since, upon change, almost all intermediate
steps are visited, and, thus, the system can interact with whatever
thing that resides at those steps), that is, the \emph{classical picture
of time}. The quantum system at an event is surrounded by a dispersion
cloud of events which can visit next, and the classical proper time
is just some average $<\tau>$ of that dispersion, and the actual
quantum transitions measure how much the actual quantum time \emph{deviates}
from this average $<\tau>$.

$\;$

But this means that quantum collapse modeled as some explicit time
evolution or dynamics is untenable here, and that only the remaining
one is consistent. Indeed, in our view, the standard classical time,
which, among other things, is used as the external time parameter
to define, e.g., the Schrödinger time evolution or any modification,
is only an emergent feature at the macro level, and fuelled at the
micro, fundamental level by an irreducible collapse. Thus, the time
evolution under some classical time parameter $\tau\approx<\tau>$
will \emph{break} when something from the fundamental level \emph{leaks}
to the macroscopic level. And, of course, this is precisely the situation
in a quantum measurement, when the state collapses when certain two
systems interact: the classical time will never be able to explain
this process since the latter intervenes precisely in making possible
the quantum, and therefore also the classical, time. 

$\;$

In classical GR, time is a layered notion consisting in \cite{key-3}:
i) mutability; ii) duration; iii) change. Only the first two are explicitly
modeled by GR's formalism, and its results are read as those for a
changing world, but change remains a baffling thing that has to be
taken as ontologically fundamental and without further explanation.
But now, in the previous QG possible scenario and unlike the classical
case, we don't need to introduce change in an ad-hoc manner here,
since change can be seen as arising from quantum collapse (\emph{now
taken as ontologically fundamental} \emph{and irreducible}) after
an interaction. We take the collapse as the \emph{only} source of
change and actually \emph{identify} it with it (in other views, collapse,
of course, implies change, but the converse is not necessary; here
we say they are indeed the same thing.) Thus, the change in the classical
theory actually comes from the fundamental quantum theory, of which
the former is a limit. Furthermore, in light of this, then the argument
used to show the necessity of the collapse\footnote{The argument consists in noting that, due to the Kochen-Specker theorem,
not all values can be defined simultaneously for a same state, it's
only possible locally for any maximal set of commuting properties;
thus, if the value of a property $A$ is defined in an initial state
as well as that of another property $C$, which commutes with the
first one but also with another property $B$ (whose value is undefined
in the initial state) which is non-commuting with $A$, a state that
makes $B$ to have a value cannot be the same as the initial one,
since $A$ cannot have a defined value in this second state due to
the non-commutation, and, in addition, the value of $C$ in the second
state may be different from the first value. In this way, it's clear
that the passing from the initial state to the second one will involve
a jump in the definition/indefinition of the values of $A$ and $B$,
and a possible jump in the value of $C$. But this doesn't say if
this jump actually ever happens in reality. Now, experimentally, it's
known that certain type of interactions select a state for the system.
Thus, if we subject the system at the same time to two interactions,
the first which selects the previous initial state and the second
which selects the other state, then we reach a paradox since those
states cannot be the same. What this means, of course, is that we
cannot have at the same time the two previous interactions happening
to the system. One solution is that only one of the interactions happens
in the present universe and then the other is forever forbidden to
happen to the system. A second one is that a particular type of change
given by the mentioned jump does exist and that one interaction can
only happen after the other in this case (in our approach, this type
of change is, furthermore, the only one that actually exists.) Of
course, given the experimental results, it's clear that the correct
option is the second one. What's interesting of this is that one is
forced to postulate it by the quantum formalism itself (although,
supplemented with some experimental inputs, as we saw.) In classical
physics, where all values are always defined, no paradox is present,
and then one can add or subtract the notion of change without producing
any logical contradiction inside the formalism. We note that quantum
collapse cannot be ``illusory'', since its necessity comes from
very basic feautures of quantum theory. This means that, in our approach,
change cannot be ``illusory'' either, then.} in QM can be now used to show the necessity of change, which then
becomes a quantum phenomena and very tied to the characteristic non-commutativity
of quantum properties.

$\;$

Thus, QG may achieve an \emph{ontological synthesis} in the ontology
of today's physics. At one hand, QM informs GR about the quantization
of proper time, on the other, GR, by puting time itself as something
to quantize, brings its layers, change in particular, to inform QM.

$\;$

\end{document}